\documentclass{article}
\usepackage{spconf,graphicx}
\usepackage{amsmath,amssymb,amsfonts,mathrsfs}
\usepackage{algorithmic}
\usepackage{array}
\usepackage{textcomp}
\usepackage{stfloats}
\usepackage{verbatim}
\usepackage{graphicx}
\usepackage{booktabs}
\usepackage{multirow}
\usepackage{xcolor}
\usepackage{etoolbox,siunitx}
\usepackage{xurl}
\usepackage{hyperref}
\usepackage{threeparttable}
\usepackage{soul}

\usepackage{setspace}

\renewenvironment{thebibliography}[1]{
  \begin{oldthebibliography}{#1}
}
{
  \end{oldthebibliography}
}
\usepackage[
backend=biber,
style=ieee,
citestyle=numeric-comp,
maxbibnames=3,
maxcitenames=3,
doi=false,isbn=false,url=false,eprint=false,
]{biblatex}
\defbibheading{bibliography}[\refname]{}

\title{Mamba-based Decoder-Only Approach with Bidirectional \\ Speech Modeling for Speech Recognition
}
\name{Yoshiki Masuyama$^{1}$, Koichi Miyazaki$^{2}$, Masato Murata$^{2}$}
\address{$^1$Tokyo Metropolitan University, $^2$CyberAgent, Inc., Japan}
\begin{document}
\ninept
\maketitle
\setlength\abovedisplayskip{1.5mm}
\setlength\belowdisplayskip{1.5mm}
\begin{abstract}
Selective state space models (SSMs) represented by Mamba have demonstrated their computational efficiency and promising outcomes in various tasks, including automatic speech recognition (ASR).
Mamba has been applied to ASR task with the attention-based encoder-decoder framework, where the cross-attention mechanism between encoder and decoder remains.
This paper explores the capability of Mamba as the decoder-only architecture in ASR task.
Our MAmba-based DEcoder-ONly approach (MADEON) consists of a single decoder that takes speech tokens as a condition and predicts text tokens in an autoregressive manner.
To enhance MADEON, we further propose speech prefixing that performs bidirectional processing on speech tokens, which enriches the contextual information in the hidden states.
Our experiments show that MADEON significantly outperforms a non-selective SSM.
The combination of speech prefixing and the recently proposed Mamba-2 yields comparable performance to Transformer-based models on large datasets.
\end{abstract}
\begin{keywords}
State-space model, Mamba, speech recognition, decoder-only model, prefix language model
\end{keywords}

\section{Introduction}
\label{sec:intro}

Transformer~\cite{transformer} and its variants~\cite{conformer,conformer-vs-e-brachformer} have dramatically improved the performance of a wide range of speech processing tasks, including automatic speech recognition (ASR).
The key to their success is the attention mechanism that can dynamically aggregate the information from the entire sequence.
Meanwhile, the attention mechanism typically suffers from its quadratic computational complexity with respect to the sequence length.
To mitigate this issue, deep state space models (SSMs) have been developed~\cite{s4,s4d,h3}.
SSMs can be trained with a sub-quadratic complexity owing to tailored algorithms, and their recurrent nature reduces the required memory during inference.
Furthermore, SSMs have shown promising performance in various speech processing tasks such as ASR~\cite{dssformer,s4decoder,mssm,s4former}, speech synthesis~\cite{sashimi}, and speech enhancement~\cite{s4m,s4ndunet}.

Existing SSMs, e.g., structured SSM (S4)~\cite{s4}, are built on linear time-invariant (LTI) systems, and their parameters are fixed regardless of the input sequence.
This input-independent architecture inhibits the capability of SSMs.
The selective SSM introduced in Mamba~\cite{mamba} dynamically computes the SSM parameters based on the input sequence and has demonstrated outstanding performance in computer vision~\cite{vision-mamba}, natural language processing~\cite{mambabyte}, and speech processing tasks~\cite{mambase,mambass,mambainspeech,miyazaki2024interspeech}.
In particular to ASR task, Mamba has been validated on the encoder-only approach with the connectionist temporal classification (CTC)~\cite{miyazaki2024interspeech} and on the attention-based encoder-decoder (AED) approach~\cite{mambainspeech,miyazaki2024interspeech}.
Notably, Mamba outperforms Transformer and S4 when used as a decoder in the joint CTC/AED framework~\cite{joint-ctc-att-decoding}.

While Mamba has been used non-autoregressively in speech applications, the decoder-only model is simple yet effective for sequence-to-sequence tasks, where the model autoregressively predicts the next token~\cite{gpt,gpt3}.
It has been successfully applied to unified speech and text processing, either by adapting a pre-trained large language model~\cite{wavprompt,asru2023decoderonly,icassp2024udagawa,iclr2024llm,qwenaudio} or by training a model from scratch~\cite{icassp2024lossmasking,arxiv2023tsunoo,viola,audiopalm,voxtlm}.
Most of these models are based on Transformer and require quadratic complexity to handle a long sequence comprising speech and text tokens.
On the other hand, Mamba can reduce the computational complexity, while it has shown promising performance as a decoder in the joint CTC/AED framework~\cite{miyazaki2024interspeech}.

\begin{figure}[t]
    \centering
    \includegraphics[width=0.75\linewidth]{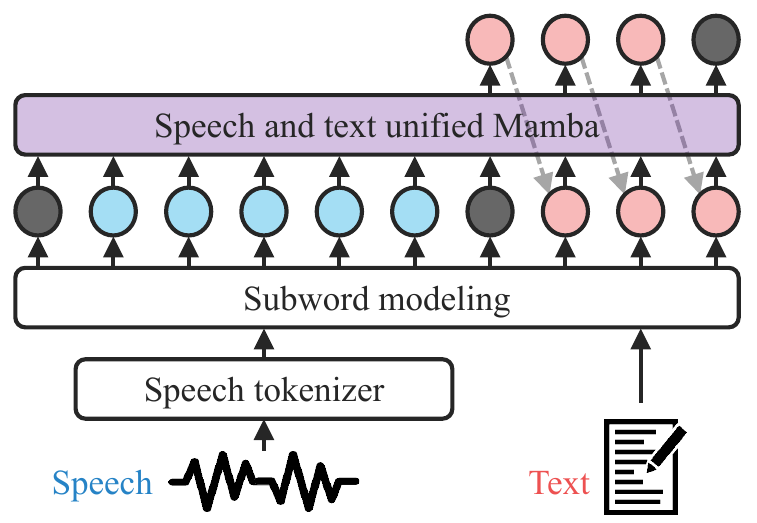}
    \vskip -0.1in
    \caption{Overview of MADEON for ASR task.
    The blue and red circles show the speech and text tokens obtained through subword modeling, respectively.
    The black circles are special tokens, and the gray dotted lines indicate the autoregressive text generation.
    }
    \label{fig:overview}
    \vskip -0.15in
\end{figure}

In this paper, we explore a MAmba-based DEcoder-ONly approach (MADEON) in ASR task towards SSM-based unified speech and text modeling.
As depicted in Fig.~\ref{fig:overview}, MADEON employs a single Mamba decoder that takes speech tokens as a condition and predicts the transcription in an autoregressive manner.
We further propose speech prefixing, which performs bidirectional processing on speech tokens to enhance the contextual modeling capability of MADEON.
We also investigate Mamba-2~\cite{mamba2} that can leverage larger hidden states more efficiently than the original Mamba. 
Our experiments show that Mamba significantly improves the word error rate (WER) from a non-selective SSM.
Although the unidirectional MADEON lags behind Transformer-based models, the integration of speech prefixing and Mamba-2, MADEON-2SP, achieves a comparable performance to Transformer-based models on large datasets.
Our contributions are summarized as follows:
\begin{itemize}
    \item We explored the efficacy of Mamba in a decoder-only approach while existing studies with Mamba were built upon the AED approach~\cite{mambainspeech,miyazaki2024interspeech}.
    \item We proposed speech prefixing to enhance the contextual modeling capability of MADEON.
    \item We confirmed the effectiveness of Mamba-2 in ASR task.
\end{itemize}

\section{Related Works}

\subsection{Overview of S4 and Mamba}

SSMs have gained much attention as an alternative to recurrent neural networks and Transformers due to their efficiency in capturing long-range dependencies~\cite{s4,s4d,h3}.
SSMs are typically based on LTI systems and map a sequence $\mathbf{x}_l \in \mathbb{R}^M$ to $\mathbf{y}_l \in \mathbb{R}^M$ by leveraging hidden states.
For instance, a time-invariant SSM handles each entry of $\mathbf{x}_l$ and $\mathbf{y}_l$ separately, and its discretized formulation is given by
\begin{subequations}\begin{align}
    \mathbf{h}_{m,l} &= \overline{\mathbf{A}}_{m} \mathbf{h}_{m,l-1} + \overline{\mathbf{b}}_m x_{m,l},
    \label{eq:recurrent} \\
    y_{m, l} &= \mathbf{c}^\mathsf{T} \mathbf{h}_{m, l} +  d_m x_{m, l}, \\
    \overline{\mathbf{A}}_m, \overline{\mathbf{b}}_m &= \exp (\Delta_m \mathbf{A}), \Delta_m \mathbf{b}, \label{eq:discretize}
\end{align} \label{eq:ssm}%
\end{subequations}
where $\mathbf{h}_{m, l} \in \mathbb{R}^N$ is the hidden state for the $m$-th entry of the features, and $(\cdot)^\mathsf{T}$ denotes the transpose.
The SSM parameters, $\mathbf{A} \in \mathbb{R}^{N \times N}$, $\mathbf{b} \in \mathbb{R}^{N}$, $\mathbf{c} \in \mathbb{R}^{N}$, and $d_m \in \mathbb{R}$ are optimized together with other parameters of a neural network.
In \eqref{eq:discretize},  $\Delta_m \in \mathbb{R}_+$ represents the time step for discretizing $(\mathbf{A}, \mathbf{b})$.
Despite its recurrent nature in \eqref{eq:recurrent}, we can train SSM in sequence parallel by using a structured matrix for $\mathbf{A}$~\cite{s4}.
This paper assumes its diagonality.

Typical SSMs, e.g., S4~\cite{s4}, are not designed for input-dependent processing.
Mamba introduces a selection mechanism that computes the SSM parameters from the input sequence~\cite{mamba}:
\begin{subequations}\begin{align}
    \mathbf{b}_l, \mathbf{c}_l &= \mathbf{W}_B \mathbf{x}_l, \mathbf{W}_C \mathbf{x}_l, \\
    \Delta_{m,l} &= \texttt{softplus}(\Delta_m + \mathbf{w}_{\Delta}^\mathsf{T} \mathbf{x}_l),
\end{align} \label{eq:mamba}%
\end{subequations}
where $\mathbf{W}_B \in \mathbb{R}^{N\times M}$, $\mathbf{W}_C \in \mathbb{R}^{N\times M}$, and $\mathbf{w}_\Delta \in \mathbb{R}^{M}$ are the additional parameters of the neural network, and  $\texttt{softplus}(\cdot)$ refers to $\log(1+\exp(\cdot))$.
By replacing the time-invariant parameters in \eqref{eq:ssm} by $(\mathbf{b}_l, \mathbf{c}_l, \Delta_{m,l})$, Mamba outperforms various non-selective SSMs~\cite{mamba}.
Although the efficient algorithm used in S4 is not applicable, its training leverages the parallel scan~\cite{s5} to avoid sequential recursion and reduces the memory requirement by  recomputation.

Mamba has been applied to various speech processing tasks such as ASR~\cite{mambainspeech,miyazaki2024interspeech}, speech synthesis~\cite{miyazaki2024interspeech}, and speech enhancement~\cite{mambase,mambainspeech}.
These studies focus on the efficiency of Mamba, and Mamba is used to handle an entire sequence non-autoregressively.
A paper relevant to ours~\cite{miyazaki2024interspeech} uses Mamba in the decoder of the joint CTC/AED-based framework~\cite{joint-ctc-att-decoding}.
It demonstrates the benefit of Mamba in the decoder but still requires the cross-attention mechanism between the encoder and decoder.
Meanwhile, we explore the efficacy of Mamba in an attention-free decoder-only model.

\subsection{ASR with discrete speech tokens}

Discrete speech tokens are a compact alternative representation to high-dimensional real-valued features~\cite{ann2022acl,audiolm,yifan2024icassp} and suitable for unified speech and text modeling~\cite{audiopalm,voxtlm}.
Semantic tokens, obtained by $k$-means clustering on self-supervised learning (SSL) features, have shown superior ASR performance to discrete tokens obtained by other techniques~\cite{ASR2}.

During $k$-means clustering, the cluster centers $\{\boldsymbol{\mu}_1, \ldots, \boldsymbol{\mu}_K\}$ are optimized on a training dataset, where $K$ is the number of clusters.
The discrete tokens for each utterance $(k_1, \ldots, k_T)$ are obtained by assigning a cluster index to the SSL features $(\mathbf{z}_1, \ldots, \mathbf{z}_T)$:
\begin{equation}
    k_t = \arg\min_k \| \mathbf{z}_t - \boldsymbol{\mu}_k \|_2^2,
    \label{eq:vq}
\end{equation}
where $t = 1, \ldots, T$ denotes the frame index.

The sequence of the cluster indices typically contain repetition and co-occurrences.
To reduce the redundancy, previous studies remove the repetition and apply subword modeling~\cite{ASR2}:
\begin{align}
    \mathcal{O} &= (o_1, \ldots, o_{L_\text{speech}}) \nonumber \\
    &= \texttt{Subwording}(\texttt{DeDuplication}(k_1, \ldots, k_T)),
    \label{eq:asr2}
\end{align}
where $L_\text{speech}$ is the number of discrete speech tokens after subword modeling.
These tokens $\mathcal{O}$ are passed to a neural network along with text tokens.
We compute the discrete speech tokens via ESPnet~\cite{espnet} and use SentencePiece~\cite{sentencepiece} for subword modeling.

\section{Proposed method}
\label{sec:proposal}

In this section, we present the Mamba-based decoder-only approach (MADEON) as depicted in Fig.~\ref{fig:overview}.
Furthermore, we introduce speech prefixing to enhance its performance.

\subsection{Unidirectional MADEON for ASR task}

Let $\mathcal{W} = (w_1, \ldots, w_{L_\text{text}})$ be the text sequence representing the transcription, where $L_\text{text}$ is the number of text tokens after subword modeling.
To predict $\mathcal{W}$ from the discrete speech tokens $\mathcal{O}$, MADEON performs the next token prediction for the text tokens while taking the discrete speech tokens as a condition:
\begin{align}
    p(\mathcal{W})
    &= \prod_{l=1}^{L_\text{text}+1} p(w_l \mid w_0, w_1, \ldots, w_{l-1}, \mathcal{O}) \nonumber \\
    &= \prod_{l=1}^{L_\text{text}+1} \texttt{MADEON}(w_0, w_1, \ldots, w_{l-1}, \mathcal{O}),
    \label{eq:prefixlm}
\end{align}
where $w_0$ and $w_{L_\text{text}+1}$ are special tokes, \texttt{<BOS>} and \texttt{<EOS>}, respectively.
We further add another special token indicating the beginning of speech, \texttt{<Speech>}, to $\mathcal{O}$ as $o_0$.

\begin{figure*}[t]
    \centering
    \includegraphics[width=0.98\linewidth]{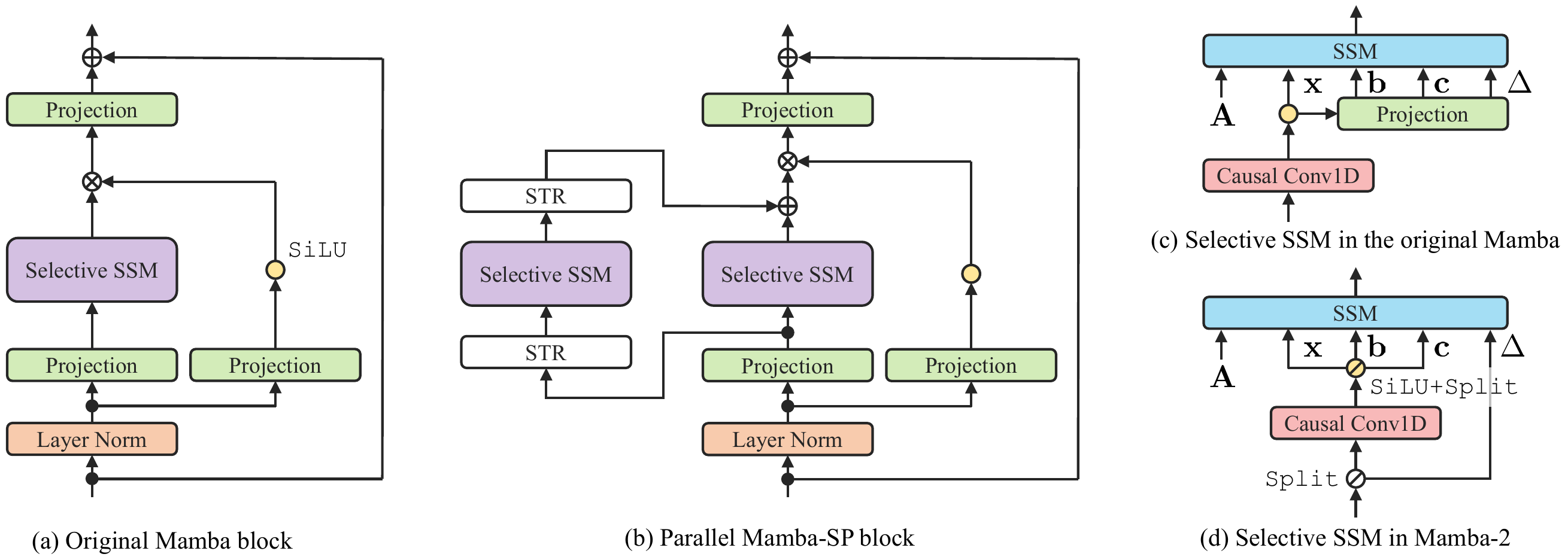}
    \vskip -0.1in
    \caption{
        Architecture of (a) the original Mamba block and (b) the parallel Mamba-SP block.
        The selective SSM blocks used in the original Mamba and Mamba-2 are shown in (c) and (d), respectively.
        The symbol $\oslash$ indicates that a single vector is split into multiple vectors~\cite{mamba2}.
        STR denotes the speech token reversal whose detail is shown in Fig.~\ref{fig:str}.
    }
    \label{fig:mambas}
    \vskip -0.15in
\end{figure*}

MADEON consists of an embedding layer, a series of Mamba blocks, and an output layer.
The embedding layer converts the discrete tokens into $M_\text{in}$-dimensional embeddings, and the output layer predicts the next token.
The architecture of the Mamba block follows the original implementation~\cite{mamba} as depicted in Fig.~\ref{fig:mambas} (a).
Within the Mamba block, the input feature is expanded to $\mathbb{R}^M$ via an input projection layer.
The selective SSM block mixes the information across tokens, where SSM uses the input-dependent parameters $(\mathbf{b}_l, \mathbf{c}_l, \Delta_{m,l})$ given by \eqref{eq:mamba}.
An output projection layer converts the features back to $\mathbb{R}^{M_\text{in}}$.
During inference, Mamba can leverage the hidden states in its recurrent formulation \eqref{eq:ssm}.
With the hidden states of all the Mamba blocks $\mathbf{H}_{l-1}$, we can reformulate \eqref{eq:prefixlm} as follows:
\begin{equation}
    p(\mathcal{W}) = \prod_{l=1}^{L_\text{text}+1} \texttt{MADEON}(w_{l-1}, \mathbf{H}_{l-1}),
    \label{eq:autoregressive}
\end{equation}
which enables efficient inference.
Furthermore, the training of MADEON requires only subquadratic complexity with respect to the sequence length due to parallel scan.
The cross-entropy loss is computed only on the text tokens with teacher forcing.

\subsection{MADEON with speech prefixing (MADEON-SP)}

\begin{figure}[t]
    \centering
    \includegraphics[width=0.8\linewidth]{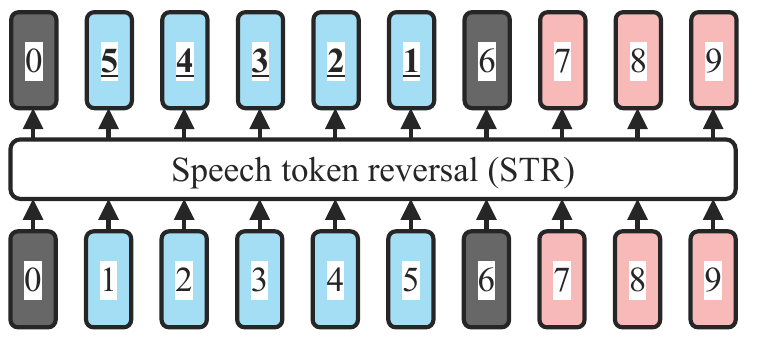}
    \vskip -0.1in
    \caption{Illustration of the speech token reversal that rearranges the features of speech tokens in reverse order.
    Features of speech and text tokens are colored by blue and red, respectively.}
    \label{fig:str}
    \vskip -0.15in
\end{figure}

In MADEON, Mamba performs unidirectional processing for both speech and text tokens.
Meanwhile, bidirectional Mamba has shown its efficacy in an encoder of the AED framework~\cite{mambainspeech,miyazaki2024interspeech} similar to well-developed bidirectional RNNs.
However, bidirectional Mamba is not directly applicable to an autoregressive decoder-only model.
We, thus, propose MADEON with speech prefixing (MADEON-SP) that performs bidirectional processing only on speech tokens while preserving unidirectional processing for text tokens.
We expect that speech prefixing enriches the contextual information in the hidden states through bidirectional speech modeling.
To realize MADEON-SP, we introduce a speech token reversal that rearranges the features of speech tokens in reverse order, as depicted in Fig.~\ref{fig:str}, and design two variants of the Mamba block.

\noindent\textbf{Parallel Mamba-SP block:}
Fig.~\ref{fig:mambas} (b) shows the parallel Mamba-SP block inspired by the vision Mamba~\cite{vision-mamba}.
This architecture shares the layer normalization and projection layers for both forward and backward modeling, enabling efficient bidirectional processing.
By applying the speech token reversal before and after the backward selective SSM, we preserve the original temporal order of the tokens.

\noindent\textbf{Serial Mamba-SP block:}
A serial Mamba-SP block alternately stacks the original unidirectional Mamba block and the speech token reversal inspired by \cite{mamband}.
The Mamba block following the speech token reversal performs backward modeling for speech tokens, where the parameters are not shared with the forward processing.

Speech prefixing is closely related to prefix language modeling (prefixLM) that allows a decoder-only model to leverage bidirectional context within a condition~\cite{t5}.
In the case of Transformer-based models, prefixLM is realized by amending the attention mask to allow non-causal attention within the prefix.
SSMs are inherently unidirectional, and thus we introduce the speech token reversal.

\subsection{MADEON-2 based on Mamba-2}

Mamba-2 is another selective SSM that incorporates the multihead patterns inspired by Transformer and simplifies $\mathbf{A}_m$ to a scalar~\cite{mamba2}.
These modifications allow to increase the state size $N$ with a moderate number of parameters and to derive a hardware-efficient algorithm.
It is advantageous to increase the state size because MADEON should preserve the speech context in the hidden states.
In Mamba-2, the input feature $\mathbf{x}_l \in \mathbb{R}^M$ is reshaped into $I$ heads of dimension $J$, where $IJ = M$.
The scalar SSM for Mamba-2 is defined as follows:
\begin{subequations}\begin{align}
    \mathbf{h}_{i,j,l} &= \overline{a}_{i,l} \mathbf{h}_{i,j,l} + \overline{\mathbf{b}}_{i,l} x_{i,j,l}, \\
    y_{i, j, l} &= \mathbf{c}_l^\mathsf{T} \mathbf{h}_{i,j,l} +  d_i x_{i,j,l}, \\
    \overline{a}_{i, l}, \overline{\mathbf{b}}_{i, l} &= \exp (\Delta_{i,l} a_i), \Delta_{i, l} \mathbf{b}_l,
\end{align} \label{eq:sssm}%
\end{subequations}
where $i = 1, \ldots, I$ is the head index, $j = 1, \ldots, J$ is the index in each head, and the SSM parameters are shared within each head.
In contrast to the original Mamba, Mamba-2 computes the SSM parameters in parallel with the input feature $\mathbf{x}_l$, as illustrated in Fig. ~\ref{fig:mambas} (d)%
\footnote{
We opt not to use the extra normalization layer introduced in \cite{mamba2} due to instabilities in our preliminary experiments.
}, which reduces the number of parameters.
We develop MADEON-2 by replacing Mamba in MADEON with Mamba-2.
Since the difference between Mamba and Mamba-2 is the design of the selective SSM blocks as shown in Fig.~\ref{fig:mambas} (c)--(d), speech prefixing is easily incorporated with MADEON-2 as MADEON-2SP.

\section{Effectiveness of speech prefixing}
\label{sec:exp1}

\subsection{Experimental setups}

We first investigate the ASR performance of the decoder-only approach with different SSMs and demonstrate the benefit of speech prefixing.
We used the ESPnet~\cite{espnet} for training and evaluation%
\footnote{
Our configurations and training scripts are available online: \url{https://github.com/YoshikiMas/madeon-asr}.
}.

\noindent\textbf{Data and pre-processing:}
We used the 100h subset of the LibriSpeech dataset~\cite{librispeech-corpus}.
Following \cite{ASR2}, we augmented the training data with speed perturbation of factors 0.9 and 1.1 and used the WavLM~\cite{wavlm}%
\footnote{\url{https://huggingface.co/microsoft/wavlm-large}}
features of the 21st layer for $k$-means clustering.
The number of clusters $K$ was set to 2,000.
We performed de-duplication and subword modeling as in \eqref{eq:asr2} with 10,000 subword units.

\noindent\textbf{Models:}
MADEON consisted of the 16 Mamba blocks, where the embedding dimension $M_{\text{in}} = 384$, the Mamba input dimension $M = 1536$, and the state size $N = 16$.
When using the parallel Mamba-SP block, we reduced the state size for each direction to $8$ to align the number of parameters to the unidirectional model.
Meanwhile, the serial Mamba-SP block used the same state size, i.e., $N=16$.
As Mamba-2 can increase the state size without rapidly growing the model size, we set $N$ to 128 for both unidirectional and bidirectional cases, where the head dimension $J$ was 64.

\noindent\textbf{Training:}
The AdamW optimizer with 5,000 warm-up steps was used with the peak learning rate at $0.006$.
We randomly masked out input token embeddings~\cite{icassp2024lossmasking}.
The training of MADEON-2SP took about one day with a single A100 GPU.

\begin{table}[t]
  \caption{WER (\%) for different SSMs on LibriSpeech 100h.
  Params refers to the total number of parameters ($\times 10^6)$.}
  \label{tab:ssms}
  \centering
  \resizebox {\linewidth} {!} {
  \begin{tabular}{ccc|cccc}
    \toprule
    \multicolumn{3}{c|}{Model} & \multicolumn{2}{c}{Dev WER (\%)} & \multicolumn{2}{c}{Test WER (\%)}\\
    SSM & Prefix & Params & clean & other & clean & other \\
    \midrule
    S4 & - & 32.9 & 39.8 & 39.3 & 39.8 & 40.1 \\
    Mamba & - & 38.5 & 4.9 & 7.4 & 5.0 & 8.3 \\
    Mamba-2 & - & 37.9 & 4.7 & 7.5 & 4.7 & 8.2 \\
    \midrule
    Mamba & serial & 38.5 & 4.3 & 7.0 & 4.3 & 7.5  \\
    Mamba & parallel & 39.9 & 4.4 & \bf{6.8} & 4.4 & 7.4 \\
    Mamba-2 & serial & 37.9 & \bf{4.2} & 6.9 & 4.3 & 7.6  \\
    Mamba-2 & parallel & 38.0 & 4.3 & \bf{6.8} & \bf{4.2} & \bf{7.3} \\
    \bottomrule
  \end{tabular}
  }
  \vskip -0.1in
\end{table}

\subsection{Results}

Table~\ref{tab:ssms} compares WER of different SSMs.
Among the unidirectional SSMs, Mamba significantly outperformed a non-selective SSM, S4.
Intuitively, the decoder-only approach in the ASR task is relevant to the selective copying task~\cite{mamba} that aims to output some specified tokens in an input sequence.
This task requires selectively remembering or ignoring the input tokens, and S4 fails while Mamba achieves almost 100\% accuracy~\cite{mamba}.
Since the decoder-only approach also requires selectively remembering the speech tokens, we expect Mamba to be essential.
Mamba-2 moderately improved WER by increasing the state size with the scalar SSM given by \eqref{eq:sssm}.

\begin{figure}[t]
    \centering
    \includegraphics[width=0.99\linewidth]{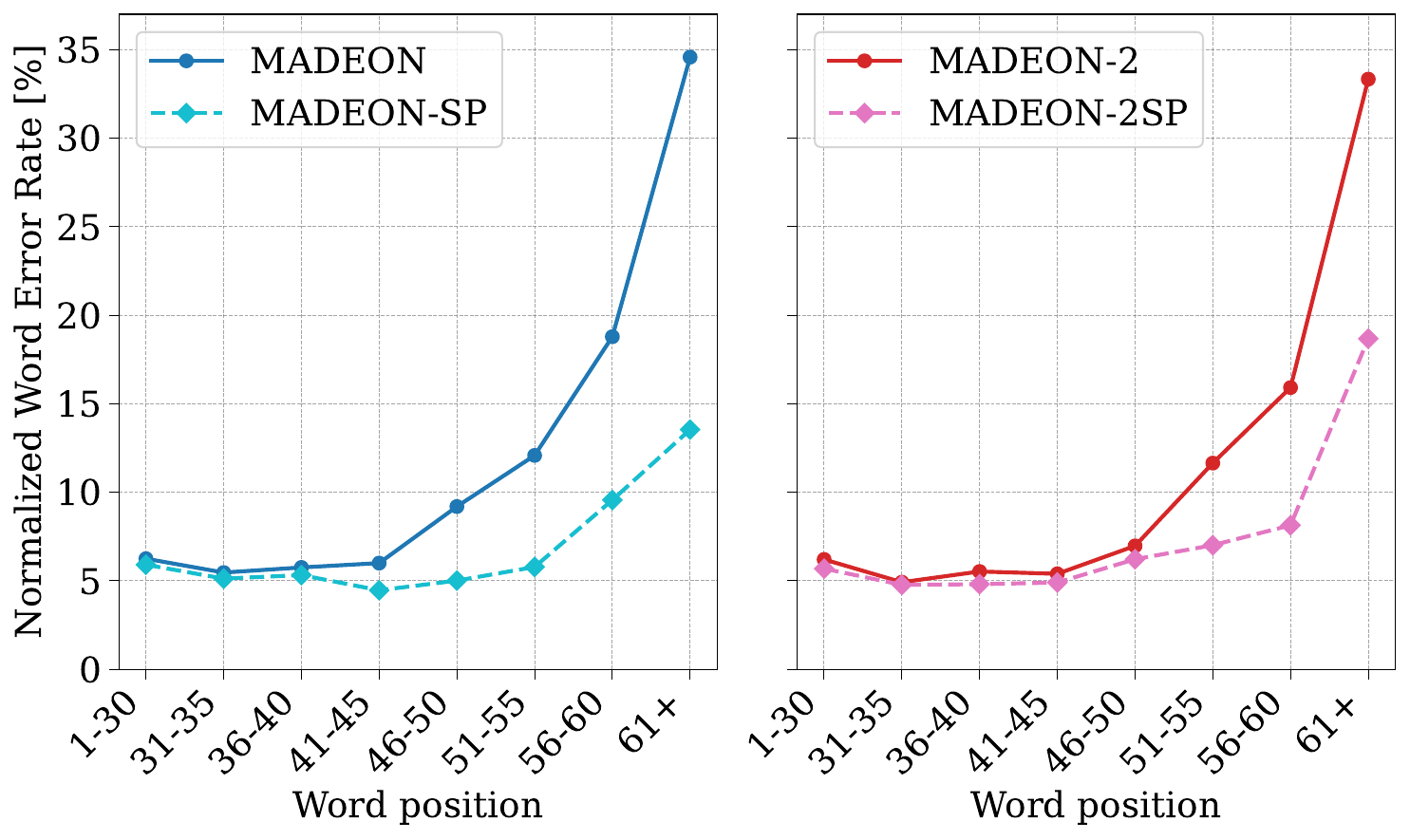}
    \vskip -0.1in
     \caption{Illustration of normalized WERs of MADEON and MADEON-2 with and without speech prefixing across different word positions on LibriSpeech 100h.}
    \label{fig:deletion_error}
\end{figure}

MADEON-SP with both serial and parallel configurations improved WER compared to the unidirectional MADEON.
In particular, we observed a substantial reduction in WER around the end of long-form speech.
Fig.~\ref{fig:deletion_error} depicts the normalized WERs across different word positions, where we used the parallel configuration for speech prefixing.
MADEON and MADEON-2 suffered from transcribing the latter part of utterances, while speech prefixing significantly mitigated this issue.
Hence, the bidirectional modeling of speech tokens successfully enriches the contextual information in the hidden states to improve subsequent text generation.
The combination of Mamba-2 and speech prefixing performed best, which confirms the effectiveness of the integration of Mamba-2 and speech prefixing, i.e., MADEON-2SP.

\section{Comprehensive evaluation of decoder-only approach}

\subsection{Experimental setups}

We conduct a comprehensive evaluation of decoder-only approach based on Transformer, Mamba, and Mamba-2.

\noindent\textbf{Data and pre-processing:}
We used six diverse datasets to cover various acoustic conditions:
read English speech (LibriSpeech 960h and its 100h subset~\cite{librispeech-corpus}), spontaneous English speech (TEDLIUM3~\cite{tedlium3} and GigaSpeech~\cite{gigaspeech}), and non-English speech (AISHELL~\cite{aishell-corpus} and CSJ~\cite{csj}).
For English datasets, we used the WavLM features since discrete speech tokens obtained from them have shown superior performance than other discrete speech tokens~\cite{yifan2024icassp,ASR2}.
Meanwhile, we leveraged language-dependent SSL models for non-English datasets, i.e.,
Chinese HuBERT%
\footnote{\url{https://huggingface.co/TencentGameMate/chinese-hubert-large}}
for AISHELL and Japanese HuBERT%
\footnote{\url{https://huggingface.co/rinna/japanese-hubert-large}}
for CSJ.
The number of subword units was set to 10,000 regardless of datasets, and other configuration is summarized in Table~\ref{tab:datasets}.

\begin{table}[t]
  \vskip -0.1in
  \caption{Dataset-dependent configurations.}
  \label{tab:datasets}
  \centering
  \resizebox {\linewidth} {!} {
  \begin{tabular}{c|ccc}
    \toprule
    Dataset & Language & \# of clusters & \# of Epochs \\
    \midrule
    LibriSpeech 100h & EN & 2,000 & 100 \\
    LibriSpeech 960h & EN & 2,000 & 35 \\
    TEDLIUM3 & EN & 1,000 & 35 \\
    GigaSpeech & EN & 1,000 & 20 \\
    AISHELL & CH & 2,000 & 70 \\
    CSJ & JP & 2,000 & 35 \\
    \bottomrule
  \end{tabular}
  }
  \vskip -0.1in
\end{table}

\noindent\textbf{Models:}
The Transformer-based decoder consists of 12 blocks, a 384-unit attention layer with 12 heads for each, and a 2560-unit feed-forward layer to align the number of parameters to MADEON.
We did not use positional encoding as in~\cite{voxtlm}, which performed better than the model with positional encoding in our preliminary experiment.
The configuration for MADEON followed the previous experiment, and the parallel configuration was used for MADEON-SP.

\subsection{Results}

Table~\ref{tab:main-results} summarizes the main results evaluated in WER or the character error rate (CER).
Among SSMs, MADEON-2SP achieved promising performance across a wide range of datasets.
MADEON variants performed slightly worse than the Transformer-based model on small English datasets, e.g., LibriSpeech 100h.
We observed that Mamba was prone to face an overfitting problem on the small datasets, while a similar tendency was reported in the Mamba-based joint CTC/AED framework~\cite{miyazaki2024interspeech}.
This problem was alleviated on large datasets, i.e., LibriSpeech 960h and GigaSpeech, and MADEON-2SP achieved comparable performance to the Transformer-based model.
The training of MADEON-2 took 6 hours on LibriSpeech 960h, while the Transformer model required 8 hours and consumed twice as much GPU memory.
This result confirms the efficiency of Mamba-2.
An interesting finding is that MADEON outperformed the Transformer-based model on non-English datasets even without speech prefixing.
For these datasets, we also investigated the performance of MADEON with the discrete speech tokens from a multi-lingual SSL model called XLS-R.
It results in CERs of 10.8/11.2 \% and 11.6/8.8/9.6 \% on AISHELL and CSJ, respectively.
These CERs are much worse than those with the language-dependent HuBERT in Table~\ref{tab:main-results}, which suggests the importance of appropriate SSL models in ASR with discrete tokens.

\begin{table*}[t]
  \caption{ASR results for Transformer-based and SSM-based decoder-only approaches.
  The performance is evaluated by WER for English datasets and by CER for non-English corpora.
  All results are obtained without an external language model.
  }
  \label{tab:main-results}
  \centering
  \resizebox {\linewidth} {!} {
  \begin{tabular}{c @{\hskip 2pt} c c| c c c | c}
  \toprule
  \multirow{2}{*}{Dataset} & \multicolumn{1}{c}{\multirow{2}{*}{Metric}} & \multicolumn{1}{c|}{\multirow{2}{*}{Eval sets}} & \multicolumn{4}{c}{Results $\downarrow$} \\
  & & \multicolumn{1}{c|}{} & \multicolumn{1}{c}{MADEON} & \multicolumn{1}{c}{MADEON-SP} & \multicolumn{1}{c|}{MADEON-2SP} & \multicolumn{1}{c}{Transformer} \\
  \midrule
  LibriSpeech 100h~\cite{librispeech-corpus} & WER & \{dev,test\}\_\{clean,other\} & 4.9 / 7.4 / 5.0 / 8.3 & 4.4 / 6.8 / 4.4 / 7.4 & 4.3 / 6.8 / 4.2 / 7.3 & \bf{4.0} / \bf{6.6} / \bf{3.9} / \bf{7.1} \\
  LibriSpeech 960h~\cite{librispeech-corpus} & WER & \{dev,test\}\_\{clean,other\} & 2.7 / 4.8 / 2.7 / 5.2 & 2.3 / 4.7 / 2.5 / 4.8 & \bf{2.2} / \bf{4.6} / \bf{2.4} / \bf{4.7} & 2.3 / {\bf{4.6}} / {\bf{2.4}} / 4.8 \\
  TEDLIUM3~\cite{tedlium3} & WER & dev / test & 10.7 / 9.6 & 9.7 / 9.7 & 8.9 / 8.9  & \bf{8.7} / \bf{8.7} \\
  GigaSpeech~\cite{gigaspeech}  & WER & dev / test & 11.2 / 11.3 & {\bf{11.0}} / 11.2 & \bf{11.0} / \bf{11.1} & 11.1 / \bf{11.1} \\
  AISHELL~\cite{aishell-corpus} & CER & dev / test  & 5.4 / 5.6  & \bf{4.8} / \bf{5.0} & 5.0 / 5.2 & 5.5 / 5.7 \\
  CSJ~\cite{csj} & CER & eval1 / eval2 / eval3 & 5.7 / 4.3 / 4.6 & {\bf{5.1}} / {\bf{3.7}} / 4.2 & 5.2 / \bf{3.7} / \bf{4.1} & 5.9 / 4.6 / 4.9 \\
  \bottomrule
  \end{tabular}
  }
  \vskip -0.15in
\end{table*}

\begin{table}[t]
  \caption{Comparison between AED models, decoder-only models, and their variants.
  The suffix SP for decoder-only models means the bidirectional processing for speech tokens.
  }
  \label{tab:aed-vs-dec}
  \centering
  \resizebox {\linewidth} {!} {
  \begin{tabular}{cccc|cc}
    \toprule
    \multicolumn{4}{c|}{Model} & \multicolumn{2}{c}{\multirow{2}{*}{WER (\%)}} \\
    Encoder & Decoder & CTC & Params & & \\
    \midrule
    \multicolumn{4}{c|}{LibriSpeech 960h (test set)} & clean & other \\
    \midrule
    \!\!E-Branchformer & Transformer & - & 40.4 & 2.7 & 4.6 \\
    \!\!E-Branchformer & Transformer & $\surd$ & 40.4 & 2.3 & 4.3 \\
    \!\!E-Branchformer & Mamba & -  & 38.6 & 2.6 & 5.8 \\
    \!\!E-Branchformer & Mamba & $\surd$  & 38.6 & \bf{2.1} & \bf{4.2} \\
    \midrule
    - & Transformer & - & 38.6 & 2.4 & 4.8 \\
    - & Transformer-SP & - & 38.6 & 2.4 & 4.7 \\
    - & MADEON-2SP & - & 38.0 & 2.4 & 4.7 \\
    \midrule
    \multicolumn{4}{c|}{GigaSpeech} & dev & test \\
    \midrule
    \!\!E-Branchformer & Transformer & - & 38.8 & 11.2 & 11.2 \\
    \!\!E-Branchformer & Transformer & $\surd$ & 38.8 & 11.2 & 11.2 \\
    \!\!E-Branchformer & Mamba & -  & 37.1 & 11.3 & 11.3 \\
    \!\!E-Branchformer & Mamba & $\surd$  & 37.1 & 11.2 & 11.2 \\
    \midrule
    - & Transformer & - & 38.6 & 11.1 & \bf{11.1} \\
    - & Transformer-SP & - & 38.6 & 11.1 & \bf{11.1} \\
    - & MADEON-2SP & - & 38.0 & \bf{11.0} & \bf{11.1} \\
    \bottomrule
  \end{tabular}
  }
  \vskip -0.1in
\end{table}

\section{Comparison of Joint CTC/AED and decoder-only approaches}
\label{sec:exp3}

\subsection{Experimental setups}

This experiment compares the decoder-only models with AED models.
We also investigate the performance of Transformer-based prefixLM~\cite{t5} as it is relevant to speech prefixing.

\noindent\textbf{Data and pre-processing:}
We chose LibriSpeech 960h and GigaSpeech, where the same configuration as in the previous experiments was used for discretizing the WavLM features.
For the AED models, we separately applied subword modeling to speech and text tokens because the encoder and decoder handle only speech and text tokens, respectively~\cite{ASR2}.

\noindent\textbf{Model:}
We trained AED models based on the joint CTC/AED framework~\cite{joint-ctc-att-decoding}.
We constructed an encoder from 12 E-Branchformer blocks~\cite{conformer-vs-e-brachformer}, where each block had $4$ attention heads with a feed-forward layer of $1024$ units.
We explored both Transformer and Mamba decoders, where the combination of the E-branchformer encoder and the Mamba decoder has shown the best WER among Mamba-based models~\cite{miyazaki2024interspeech}.
The Transformer decoder comprises $6$ blocks with $4$ attention heads, while the Mamba decoder also consists of $6$ blocks. 
We further investigate the performance of Transformer-based prefixLM (Transformer-SP).
Its architecture was similar to the decoder-only model, whereas we allowed non-causal attention for the speech tokens.
In addition, we used the relative positional encoding presented in \cite{t5}, because training of prefixLM failed without the positional encoding.

\noindent\textbf{Training:}
The AED models were trained using multi-task learning with the CTC loss~\cite{joint-ctc-att-decoding}, where the weight for the CTC loss was $0.3$.
We performed inference with and without CTC for a fair comparison with the decoder-only models.

\subsection{Results}

Table~\ref{tab:aed-vs-dec} shows WER of the AED and decoder-only models.
Among the AED models, inference with CTC consistently improved WER.
Comparing Transformer and Transformer-SP, the performance improvement from the bidirectional speech modeling was marginal, whereas it brought a significant gain for the Mamba-based model in Table~\ref{tab:main-results}.
Hence, bidirectional speech modeling is more beneficial for Mamba.
The joint CTC/AED inference using the Mamba decoder performed best on LibriSpeech 960h.
The CTC module uses the forward-backward algorithm during training and enforces the alignment between the features and the transcription.
MADEON variants take the speech context into account only through their hidden states and do not consider the explicit alignment between speech and text tokens.
This remains as room for improvement in future work.
Nonetheless, MADEON-2SP performed best on GigaSpeech and demonstrated its potential.

\vspace{-2pt}
\section{Conclusion}

We explored MADEON, a Mamba-based decoder-only approach, in ASR task.
Furthermore, we introduced speech prefixing that performs bidirectional speech modeling to enrich contextual information in the hidden states.
Our experiments showed the advantage of Mamba in the decoder-only approach compared to S4.
The integration of the speech prefixing and Mamba-2 resulted in the best performance among the MADEON variants and was comparable to Transformer-based models on LibriSpeech 960h and GigaSpeech.


\clearpage
\section{References}
\begingroup
\setstretch{0.86}
\setlength
\bibitemsep{0.86pt}
\printbibliography
\endgroup
\end{document}